**Title:**

The global freshwater system: Patterns and predictability of green-blue water flux partitioning

**Authors:**

Daniel Althoff[1,*], Georgia Destouni[1,**]

**Affiliations:**

[1]Department of Physical Geography, Bolin Centre for Climate Research, Stockholm University, Stockholm, Sweden

*Correspondence: daniel.althoff@natgeo.su.se

**Correspondence: georgia.destouni@natgeo.su.se

[SCIENCE FOR SOCIETY]

As the main input of freshwater on land, precipitation infiltrates the soil where some of it is stored as soil water or groundwater, while some is used by vegetation for transpiration as "green water", or contributes to the flows of groundwater, rivers and streams as "blue water". The green-blue water partitioning is key for sustaining life on land and below water, and for societal water and food security. We collected data from around the world to investigate and decipher global patterns in this water partitioning and explain how it is affected by climate and human land and water use conditions. Our analyses show large-scale emergence of vegetation priority in taking or getting (by irrigation) the green water part it needs and only what remains after that use goes to feed blue water flows. This makes blue water security for other uses vulnerable to future changes in precipitation and in vegetated land, such as for irrigated crops and forestry. In this study, we also develop and train a Machine Learning model as a tool for predictive assessment of how green-blue water partitioning may change around the world under different future climate scenarios.

**Summary:**

The partitioning of precipitation (P) water input on land between green (evapotranspiration, ET) and blue (runoff, R) water fluxes distributes the annually renewable freshwater resource among sectors and ecosystems. We decipher the worldwide pattern and key determinants of this water flux partitioning (WFP) and investigate its predictability based on a machine learning (ML) model trained and tested on data for 3,614 hydrological catchments around the world. The results show considerably higher WFP to the green (ET/P) than the blue (R/P) flux in most of the world. Land-use changes toward expanded agriculture and forestry will increase this WFP asymmetry, jeopardizing blue-water availability and making it more vulnerable to future P changes for other sectors and ecosystems. The predictive ML-model of WFP developed in this study can be used with climate model projections of P to assess future blue and green water security for various regions, sectors, and ecosystems around the world.

# Introduction

Precipitation (P) is the main input source of freshwater on land, where the water is further partitioned to blue (runoff, R) and green (evapotranspiration, ET) water fluxes[1]. This water flux partitioning (WFP) has major implications for aquatic and terrestrial ecosystems[2,3], water and food security[4,5], and different societal sectors[6,7]. The WFP is regulated by land-atmosphere interactions[1,8] and is, therefore, vulnerable to disturbances and shifts in these. Around the world, regional and continental studies have found shifts in WFP due to climate change, direct human changes in land and water use, or both[9–13]. Yet, a conclusive understanding of how WFP inherently relates to various combinations of such changes at large scale still remains to be reached. Predictive capability for WFP also needs to be improved and further developed[14,15] as climate projections suggest more warming and shifted rainfall patterns with possible more intense heatwave, rainfall, drought and flood events in the coming decades[16].

Understanding how WFP depends on changing climate and land use conditions around the globe is essential, for example, for adaptation to climate change and generally for freshwater security[17]. However, high uncertainty has been attributed to Earth System Models' ability to represent historical hydroclimatic covariation patterns and trends in both regional and global assessments[11,14,15,18,19]. It is, therefore, important to also explore additional, complementary modelling frameworks for determining WFP[20,21]. The complex non-linear interactions between land and atmosphere and the increasing amount of accessible data make machine learning (ML) a possible important tool for meeting climate-related predictability challenges[20], although its black-box-like characteristics and possible limited applicability outside a model's initial training domain are reasons for concerns[22,23].

Here, we develop and explore a global ML model and its pattern recognition and predictability capabilities for average WFP in hydrological catchments of various scales and locations with different climate, land-use and other geographically dependent conditions around the world. To train and test the model, we compiled data for 3,614 hydrological catchments (coloured fields in Figure 1A) with continuous temporal data coverage for daily runoff (limiting data component required for catchment-wise water balance closure) over at least 25 years within the period 1980-2020 (see further Supporting Information (SI) on runoff databases in Table S1). Global data are available in this period for land-use from 1992 (SI Table S2) and for climate (temperature and precipitation) over the whole period and longer (SI Table S3). To characterize the catchments and further explore the physical-hydrological reasonableness and implications of the global ML model, we also independently quantified and related ML model results to some additional hydrological catchment characteristics. These include catchment-average aridity index (see Experimental procedures (EP) and SI Figure S1), the minimum averaging time required to obtain temporally stable average WFP in each catchment (referred to as the water flux equilibration time, $T_{eq}$; see EP and SI Figure S2), and an assumed (and tested) $T_{eq}$-related indicator of average groundwater contribution to surface water runoff under zero monthly precipitation conditions ($R_{gw0}$; see EP and SI Figure S3). The hydrological variable quantifications considered the full period of data availability for each catchment (SI Table S1), while data for the period 1992-2020 were used for ML model training and testing, given the data availability conditions for the different model variables (SI Table S3).

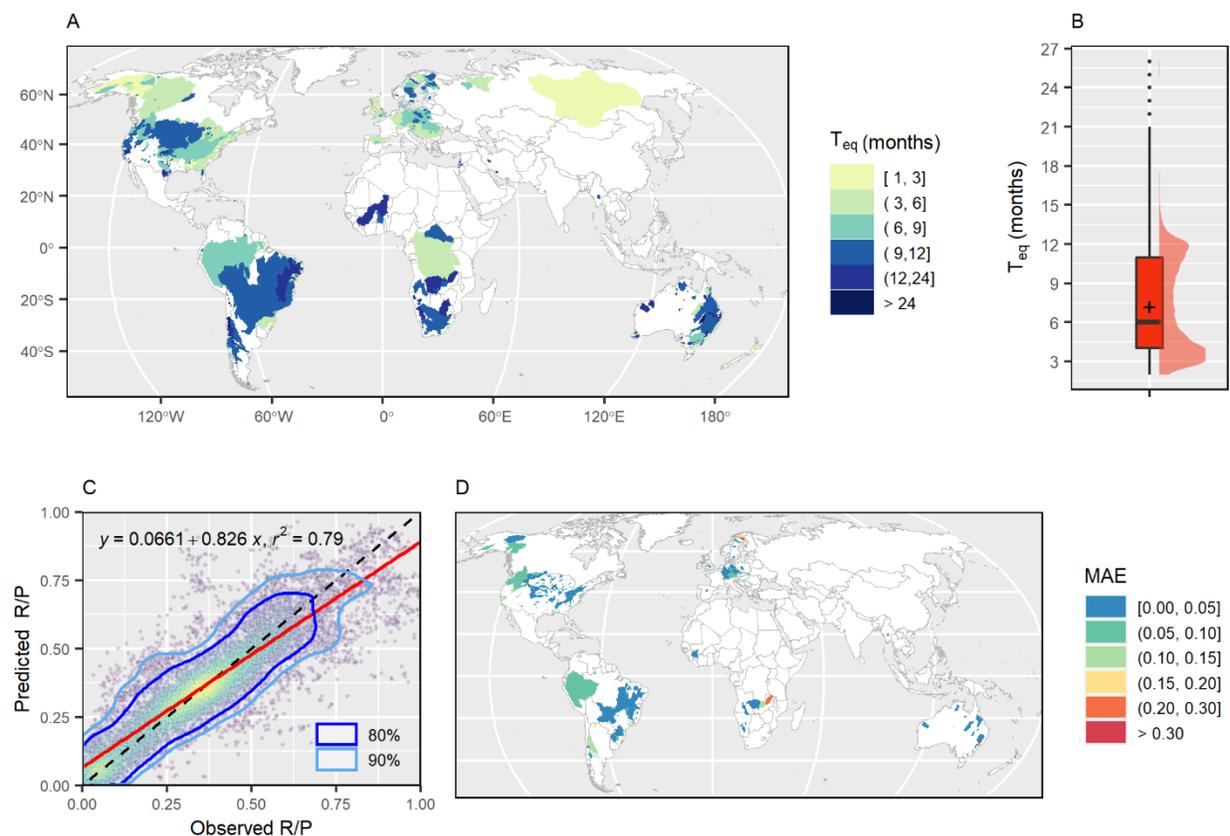

**Figure 1.** (A) Map of the 3,614 investigated hydrological catchments (colored fields) that also shows their water flux equilibration time ($T_{eq}$), along with (B) the box plot statistics of $T_{eq}$ across the catchments (with the + symbol showing the mean value). (C-D) Performance of the ML model for the blue water flux (R/P) partitioning in the test set of 757 catchments, in terms of: (C) a scatterplot of predicted versus observed R/P that also shows the location of 80% and 90% of the data points based on a two-dimensional kernel density estimation; and (D) a map of mean absolute error (MAE) obtained for each catchment.

To address general interpretability and applicability concerns with ML modelling, we adopted model interpretability techniques to decipher the key determinants of catchment-average WFP around the world and gain insight into the model decision-making. The ML model applicability to new unseen data was further investigated in relation to model variable values and independently estimated hydrological characteristics of catchments within and outside of the model area of applicability. The observational data and ML model results show that most of the world outside the polar regions has a much higher WFP to the green water flux (global area-weighted average ET/P = 0.62) than the blue water flux (global area-weighted average R/P = 0.38). For the future, the model implies that further warming, as well as land-use changes towards expansion of agricultural and forest areas, should be expected to enhance pressure on blue water availability and security, and make the blue water flux more vulnerable to future precipitation changes. The developed ML model can be combined with climate model projections to better assess these implications for different societal sectors and ecosystems around the world. Our findings also clarify and emphasize the need for additional data and ML model training, in particular for the northern subtropical and low-latitude temperate regions and the polar regions of the world.

## Results and discussion

The final database used for the present investigations includes data for the 3,614 study catchments around the world (Figure 1A) with varying temporal data coverage for the different synthesized types of data (SI Tables S1-S3). The Global Runoff Database Centre[24] (GRDC) provides required runoff data for many catchments, but a large part of these are located in Europe and North America. To further extend the total catchment number and world coverage, we also used additional runoff datasets that are openly available for catchments in some countries (United States[25], Brazil[26], Chile[27], Great Britain[28], and Australia[29]). Overall, the total set of study catchments covers a wide range of hydrological conditions, e.g., with aridity index PET/P ranging from 0.5 to 14.4 (mean: 1.2, SI Figure S1A-B) and covering the relevant ET/P versus PET/P Budyko space[30] (SI Figure S1C), where PET is potential evapotranspiration.

The independent quantification of water flux equilibration time ($T_{eq}$, SI Figure S2) yields $T_{eq}$ values of less than 2 years in ~99.5% of the study catchments (Figures 1A-B). This ensures stable average R/P and ET/P values over the selected 5-year temporal averaging window for these and the related explanatory variables in the ML model. The independent quantification of $R_{gw0}/R_{avg}$ (SI Figures S3-S4) and its emergent considerable positive correlation with $T_{eq}$ (SI Figure S5) also support consideration of $R_{gw0}/R_{avg}$ as a likely relevant groundwater indicator, since it is hydrologically realistic for $T_{eq}$ to be longer for larger relative contribution of relatively slow-flowing groundwater to total runoff (with average value $R_{avg}$) through a catchment.

For WFP pattern recognition, we trained a ML model to predict the ratio between average runoff and average precipitation for each catchment (R/P, blue WFP). From this and catchment-wise average water balance $P \approx ET + R$ (assuming small long-term average water storage change in comparison with the main water fluxes P, ET and R)[9,11,12,31,32], the corresponding average green WFP can in turn be estimated as $ET/P \approx 1 - R/P$. The ML model uses as input catchment topography and hydro-climatic indices, and land use fractions summarized over 5-year periods (see EP for details on these explanatory variables). We used 80% of the catchments to train the model and evaluated its performance on the remaining 20% (test set). For the unseen data in the test set (Figure 1C), the developed ML model was able to predict R/P with a mean absolute error (MAE) of 0.07. This should be compared to the test set average R/P value of 0.35 (and corresponding average ET/P of 0.65). The average MAE calculated for individual catchments (Figure 1D) was similar to that of the entire test set (average MAE = 0.07, median MAE = 0.04). The ML model also scored a percentage bias of 1.20% and Kling-Gupta efficiency[33] (KGE) of 0.87 for the test set.

ML model deciphering shows that, as physically expected, a greater average catchment slope implies greater average R/P and thereby lower associated ET/P in a catchment (Figure 2A). Higher average annual temperature also relates to WFP as physically expected, with higher ET/P and thereby lower R/P (Figure 2B). Changed mean annual P, however, does not imply any clear a priori expected WFP effect, as higher P could in principle lead to either maintained relative R/P and ET/P values, or increase in any of them at the expense of the other. As a main result of the ML model deciphering, we then see that the change direction (increase/decrease) of P tends to lead to a similar change direction in (increase/decrease) in R/P and thereby to an opposite change direction (decrease/increase) in ET/P (Figure 2B). This is consistent with other recent multi-catchment results for drought conditions over Europe, showing strong and fast runoff reductions in response to precipitation deficits, but relatively small/negligible decreases or even increases in some places for corresponding evapotranspiration[9].

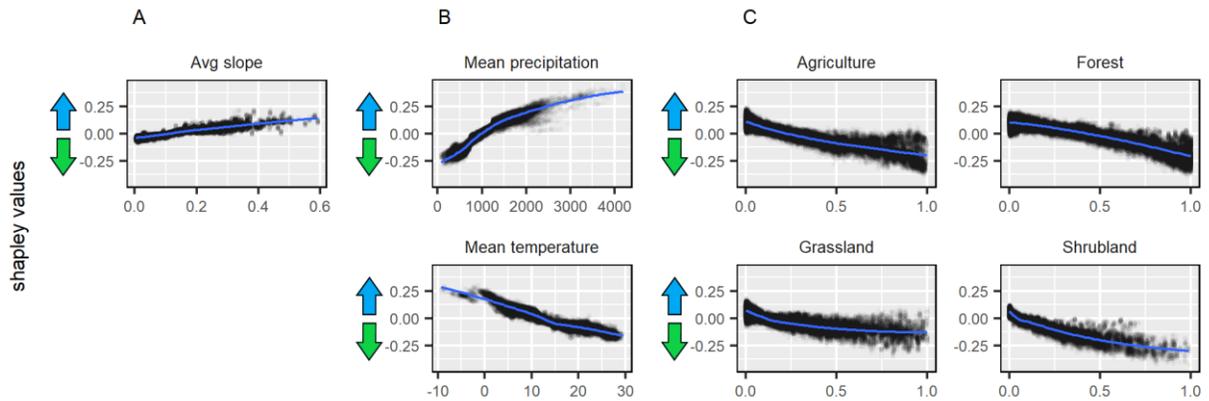

**Figure 2**. Marginal contributions (shapley values) of the main explanatory ML model variables to WFP predictions for the entire dataset. Shapley values are relative to the average dataset prediction (R/P = 0.37, ET/P = 0.63). Positive [negative] shapley values indicate that explanatory variables contribute to increase in R/P [ET/P]. Panels A-C show results for the most contributing explanatory variables of (A) topography, (B) climate (precipitation, temperature), and (C) land-use area fractions (of agriculture, forest, grassland, shrubland, relative to total catchment area). Figure S6 shows resulting shapley values for all explanatory variables in the ML model.

From ML model deciphering with regard to different types of land cover/use (Figure 2C), larger relative vegetated area (and greater vegetation density, i.e., more for forests than grasslands) tends to imply higher ET/P and thereby lower R/P. This effect is most pronounced for managed vegetation, like agriculture and forests, for which R/P exhibits near-linear and super-linear negative correlation, respectively. That is, vegetation and particularly managed crops and forests tend to take their cut of water even under low P conditions. This is consistent with results in other recent studies, showing vegetation greening associated with increased ET, and reduced soil moisture and runoff[3,34,35].

The remaining explanatory variables of the ML model show only minor marginal contributions to R/P (ET/P) prediction (SI Figure S6). Among the land cover/use types, water-covered area fractions of wetlands and surface waters also exhibit negative correlation to R/P (positive to ET/P, as should be expected since the rate of evaporation from open water surfaces is generally higher than ET rate from dry land areas), while R/P correlation is negligible for settlement and bare areas. Permanent snow and ice exhibit a slight positive correlation with R/P over the small relative area range that these cover in most catchments, which may be related to the relatively high albedo of these land covers regulating energy availability[36]. Overall, the results for these remaining land cover/use variables, however, have lower confidence because most data points extend only over small relative area ranges, close to zero. Furthermore, it is not straightforward to draw general trade-off conclusions for land cover/use exchanges, since the shapley values are not free of interaction, i.e., trade-offs may be context-dependent. For example, replacing forests or shrublands for agriculture may have different effects in mountainous or flat terrains and in arid or humid climates. For each region, however, simulations can show the implications of such exchanges. For the remaining climate variables, higher seasonal amplitude in precipitation [temperature] may imply a slightly higher R/P [ET/P]. The time lag between peak P and peak T also shows a small marginal contribution, with somewhat higher ET/P [lower R/P] if P peaks in the warm season, and lower ET/P [higher R/P] if it peaks in the cold season.

Applying the developed ML model to WFP prediction around the world (see EP for details on global extrapolation) yields an overall smaller average share of the total precipitation water input going to runoff (blue water flux; Figures 3A-B) than to evapotranspiration (green water flux; Figures 3B-C). The area-weighted spatial average values of local WFP in each pixel are R/P = 0.38 and ET/P = 0.62, while the flow volume-averaged WFP of total global P into global R and global ET yields R/P = 0.46 and ET/P = 0.54. Looking at latitudinal WFP distribution (Figure 3B), ET/P is on average much higher than R/P in the tropical and temperate zones of the northern hemisphere (1ºN to 56ºN) and tropical and subtropical zones of the southern hemisphere (3ºS to 40ºS), where most land area and its vegetation are located around the world, and where hydrological conditions tend to be more water than energy limited (SI Figure S7). Around the equator, in the tropical zone, ET/P and R/P tend to be similar, on average at around 0.50 each, while ET/P is relatively small and R/P is relatively large in the higher and lower latitude parts of the temperate zones in each hemisphere (above 56ºN and below 40ºS) and decrease/increase even more, respectively, in the polar zones. These are also the regions with the most energy-limited water conditions (low aridity index, PET/P) around the world (SI Figure S7), and the polar regions, where there is little vegetation, are the only places where R/P is on average higher than ET/P.

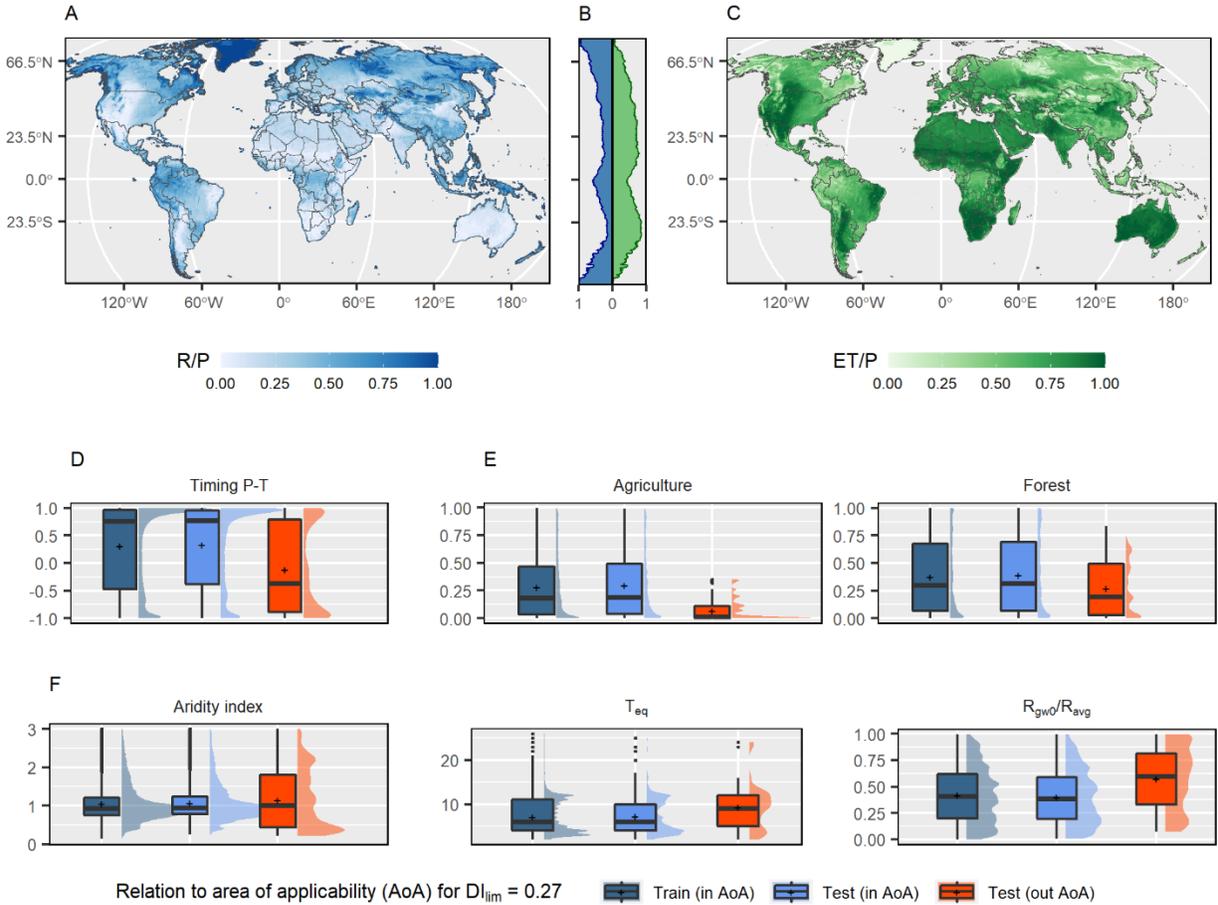

**Figure 3**. (A-C) Water flux partitioning and its average latitudinal distribution over land for (A, B-blue) R/P and (B-green, C) ET/P. (D-F) Box plot variable statistics that differ most between world parts within (training set, and part of test set) and those outside of (remaining test set part) the ML model area of applicability (AoA; considering here a dissimilarity index threshold of $DI_{lim}=0.27$, see Figures 4A-B), for the variables: (D) P-T peak timing (climatic); (E) relative agriculture and forest areas (land-use); and (F) aridity index, flux equilibration time $T_{eq}$, and indicator of groundwater runoff contribution $R_{gw0}/R_{avg}$ (model-independent hydrological characteristics). The aridity index (F) is cropped to values below 3 (2% of data not showing). SI Figure S8 shows corresponding statistics (more similar within and outside of AoA) for other explanatory variables of the ML model.

Resulting ML model error versus dissimilarity index (DI) for the study catchments suggests a relevant DI threshold in the range $0.27 \leq DI_{lim} \leq 0.50$ for classifying new unseen data as either inside or outside the model area of applicability (AoA) (Figures 3D-F and 4A-B). With regard to climate variables, the ML model tends to lose predictability for unseen data in catchments with a greater time lag between peak temperature and peak precipitation (Figure 3D). For land-use variables, predictability tends to be lost for catchments with lower relative forest and agriculture areas (Figure 3E). Looking at the model-independent hydrological catchment characteristics, predictability tends to be lost for catchments with relatively high (>> 1) or relatively low (<< 1) aridity index (implying major water or major energy limitation, respectively), and greater $T_{eq}$ and associated greater $R_{gw0}/R_{avg}$ (Figure 3F).

Figure 4A shows how the R/P (and associated ET/P) model error grows with increasing DI for the study catchments. Depending on the selection of DI threshold value ($DI_{lim}$), the ML model AoA around the world changes (Figure 4B). Selection of, for example, $DI_{lim} = 0.27$ implies that 5% of the unseen data in the test set fall outside the AoA. The resulting MAE for R/P becomes then 0.13 for the catchments outside AoA (Figure 4A), which can, e.g., be compared with the mean absolute R/P error of 0.06 for 95% of the test set data inside AoA (Figure 4A). Selecting $DI_{lim}$ in the range 0.27-0.50 does not change the error much (Figure 4A) but raises AoA from 35% to 60% of the total world land area (minus Antarctica) for increased $DI_{lim}$ from $DI_{lim1} = 0.27$ to $DI_{lim2} = 0.50$ (Figure 4B). Figure 4C maps the resulting DI for ML model application around the world and Figure 4D shows the relative world area per latitude falling within AoA for $DI_{lim1} = 0.27$ and $DI_{lim2} = 0.5$ (for spatial coverage of $DI_{lim1}$ and $DI_{lim2}$, see SI Figure S9). Overall, and regardless of specific $DI_{lim}$ choice within the range 0.27-0.50, the relative land area falling within the ML model AoA is relatively small in the northern subtropical to the low-latitude temperate region and the polar regions of the world (AoA < 50% for $DI_{lim2}$ above 66ºN, from 15ºN to 36ºN, and below 44ºS).

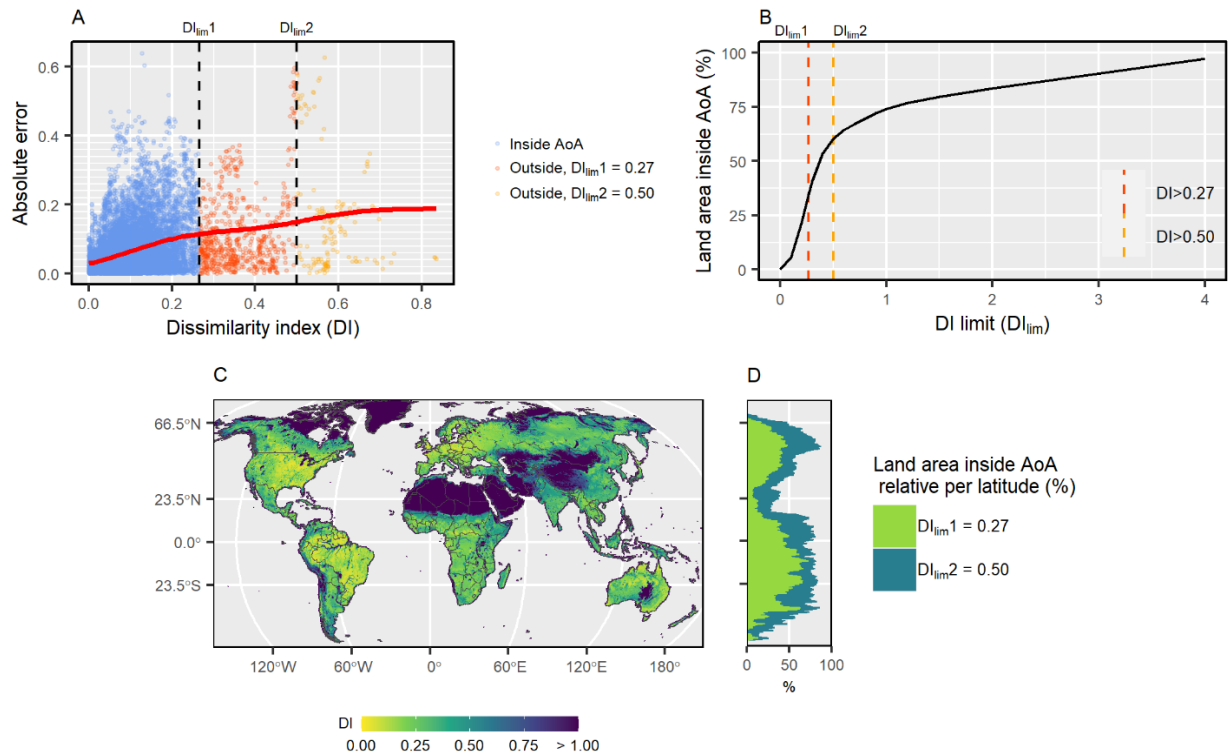

**Figure 4.** (A) Model error for the test set versus dissimilarity index (DI) for the study catchments, (B) global area of ML model applicability (AoA) for different selected DI threshold value (DI$_{lim}$) for AoA, (C) map of DI around the world, and (D) latitudinal distribution of relative land area inside AoA.

**Conclusions**

Available observational data, as well as ML model application around the world, show ET/P to be overall higher than or at lowest similar to R/P around most of the world outside the polar regions. The ML model deciphering shows that this is because vegetation, and particularly managed agricultural crops and forests predominantly take their needed water share. This is most strongly evident in the tropical and temperate zones of the northern hemisphere and the tropical and subtropical zones of the southern hemisphere, and can make blue-water availability (related to R/P) highly vulnerable to climate-driven precipitation changes and/or directly human-driven crop, forest and other vegetation changes in these regions. These world parts are also where most people live and hydrological conditions tend to be more water than energy limited, with associated relatively high water security, drought, and flood risks.

In the global ML model application for WFP prediction, around 60% [35%] of the global land area falls within the model AoA for selected DI$_{lim}$ = 0.50 [0.27]. This AoA assessment shows that additional ML model training is needed for a more accurate prediction of green-blue WFP in the northern subtropical to the low-latitude temperate region (northern Africa and most of Asia) and the polar regions of the world (north-eastern North America, Greenland, Iceland, and Russian-Arctic). In general, WFP predictability of ML trained models can be tested, deciphered and assessed for and across different hydro-climatic and land-use conditions around the world, following the model-agnostic methodology used in this study.

For the future, the ML model developed in this study implies that WFP to green fluxes (ET/P) will tend to increase to an even higher level at the expense of blue water fluxes (R/P), which will tend to get even lower, in areas where climate change drives more water limitation, and human land uses expand to more managed (and more irrigated) agriculture and/or reforestation replacing open, sparsely, or non-vegetated land. The ML model results for WFP can be combined with Earth System Model projections of future temperature and precipitation scenarios to assess corresponding blue and green water availability, security, drought and flood risks for various regions, societal sectors, and ecosystems. To enhance ML model applicability under such changes, additional data and ML model training are needed in particular for the northern subtropical and low-latitude temperate region and the polar regions of the world.

## Experimental procedures

### Resource availability

*Lead contact*

Further information and requests for resources should be directed to and will be fulfilled by the lead contact, Daniel Althoff (daniel.althoff@natgeo.su.se).

*Materials availability*

This study did not generate new unique materials.

*Data and code availability*

All input data used for the study are openly available, as stated in the article. All data generated in the study have been deposited at Zenodo and are publicly available at: https://zenodo.org/record/6519659. The code used for the machine learning model and remaining analyses has also been deposited at https://zenodo.org/record/6519659. Any additional information required to reanalyze the data reported in this paper is available from the lead contact upon request.

**Methods overview**

The methodology combines (i) an independent quantification of hydrological characteristics, and (ii) a ML modelling framework for the possible prediction of WFP worldwide. The independently quantified hydrological characteristics are the aridity index, the water flux equilibration time ($T_{eq}$), and the relative groundwater contribution to surface water runoff under zero monthly precipitation ($R_{gw0}/R_{avg}$). Besides characterizing the catchments, the quantification of $T_{eq}$ is important to ensure that the ML modelling regards temporally stable average WFP conditions for and across the study catchments with widely different hydrological conditions around the world. These hydrological characteristics are also important to explore the physical-hydrological reasonableness of the ML model applicability outside its initial domain. Below, we detail the data sources for the study, data curation, and the investigation procedures.

**Data sources**

We compiled a database for a total of 3,614 hydrological catchments around the world with available data for daily runoff (limiting data component required for catchment-wise water balance closure) over at least 25 years from 1980 to 2020. The hydrological catchments boundaries and runoff series were obtained from the Global Runoff Database Centre[24] and multiple CAMELS datasets[25–29] (for details concerning catchment selection, see SI Note S1, Table S1, and Figure S10). For these catchments, we derived daily precipitation time series from the Multi-Source Weighted-Ensemble Precipitation (MSWEP)[37] version 2.8, and daily mean temperature time series from the National Oceanic and Atmospheric Administration (NOAA) Climate Prediction Center (CPC) global temperature data[38]. Annual time series of land use/cover fractions were derived from land cover maps of the Climate Research Data Package (CRDP) that is generated within the European Space Agency (ESA) Climate Change Initiative – Land Cover (CCI-LC) project. For topography, we used the Multi-Error-Removed Improved-Terrain Digital Elevation Model (MERIT DEM)[39].

For the ML modelling framework, the target variable is the blue WFP, represented by the ratio between average runoff and average precipitation (R/P) (and the complementary, water-balanced determined green WFP – ET/P). As input, the model takes 17 explanatory variables that were calculated for each catchment concerning hydro-climatic, land use and topography indices. However, before synthesizing all of these for the catchments, we first determined a minimum time interval that ensures water flux equilibration for all catchments (see water flux equilibration $T_{eq}$ in the following section). This means that a long time series could be summarised in several data instances for a catchment. We selected a 5-year period as reference as it ensures stable flux partitioning for all catchments (Figure 1A-B). For a hypothetical observational period from 1995 to 2012, the data instances would be summarised for the sub-periods 1995-1999, 1996-2000, …, 2008-2012.

**Independently estimated hydrological characteristics**

*Aridity index*

The aridity index (AI) was calculated as the ratio between potential evapotranspiration (PET) and P (AI = PET/P). PET was calculated with the Priestley and Taylor[40] equation (Equation 1). The net radiation (Rn) was calculated following the FAO-56[41]. We used temperature data from the CPC/NOAA dataset, while relative humidity and incoming solar radiation were obtained from the National Aeronautics and Space Administration (NASA) Langley Research Center (LaRC) Prediction Of Worldwide Energy Resources (POWER) Project[42] funded through the NASA Earth Science/Applied Science Program.

$$PET = \alpha \frac{\Delta}{\Delta+\gamma} Rn \tag{1}$$

where α is an empirical constant accounting for vapor pressure deficit and resistance values, assumed as 1.26 for open bodies of water, and γ is the psychrometric constant.

*Water flux equilibration time*

To determine water flux equilibration time $T_{eq}$, we assessed the average R/P (avg(R/P)) calculated for different aggregation time scales (DT). For any DT, R/P instances are obtained from the P and R time series using a moving time-window of size DT (SI Figure S2A). These instances obtained over the whole data time series are then averaged to obtain $avg(R/P)_{DT}$. As DT increases, avg(R/P) stabilizes to a more or less constant value (SI Figure S2B). The flux equilibration time $T_{eq}$ is assumed here as the minimum DT value ($T_{eq} = DT_{min}$), for and beyond which avg(R/P) values obtained for any DT > $DT_{min}$ differ less than 1% from each other.

*Groundwater contribution to streams under zero monthly precipitation*

To estimate the relative groundwater contribution to surface water runoff under zero monthly precipitation $R_{gw0}/R_{avg}$, we regress monthly R, normalized with average monthly R over the whole runoff series ($R_{avg}$), versus monthly P for each study catchment (SI Figure S3). The groundwater indicator variable is quantified as equal to the regression line intercept (IntR) for $0 \leq IntR \leq 1$. A negative intercept (IntR < 0) implies no surface water runoff, i.e., dry streams with still-standing water, for zero monthly P. On the other hand, a positive intercept above 1 (IntR > 1) implies that the average runoff under zero monthly precipitation is larger than the total average runoff $R_{avg}$ over the whole P and R time series, which may occur for snow-dominated catchments. In both cases (IntR < 0 and IntR > 1), the groundwater flow cannot be simply estimated from the R vs P regression line. Thus, a total of 1,274 (35% of all) and 105 (3% of all) study catchments with IntR < 0 and IntR > 1, respectively, were excluded from the $R_{gw0}/R_{avg}$ analysis.

**Modelling data curation**

The modelling data are data instances corresponding to the blue WFP R/P (target) and 17 explanatory variables (input) that were summarised for every 5 years of a catchment time series. The explanatory variables refer to 5 hydroclimatic indices, 10 land use fractions, and 2 topography indices. The hydroclimatic indices were derived from the precipitation and temperature time series. These series were first approximated to sine curves to extract their corresponding annual averages, seasonal amplitude over the year, and a seasonality timing index representing the phase lag difference between peak precipitation and peak temperature over the year (for details, see SI Note S2). The ESA CCI-LC classification system describes the land uses in 38 sub-classes which were grouped into 10 broad categories that better correspond to the IPCC land categories[43] (see SI Note S3 and Table S2): agriculture, forest, grassland, shrubland, sparse vegetation, wetland, bare area, settlement, water, and permanent snow and ice. For the two topography indices, we used the MERIT DEM to derive the catchments' mean elevation and mean slope. The final database has been deposited in the study online repository[44].

**Machine learning modelling**

*Model training and testing*

For ML model training and testing, we use different parts of the total database compiled for the 3,614 hydrological catchments and the period from 1992-2020. This period was chosen based on data availability conditions for all explanatory variables (see SI Table S1 and S3) and the number of instances per catchment (average = 19.4, median = 21.0) was limited only by the length of their own runoff series. To better reflect the model prediction to new unseen areas, we used a spatial partitioning of the database, i.e., 80% of the catchments were assigned to a training set and the remaining 20% to the test set.

We chose the cubist regression for ML modelling the blue WFP R/P. The cubist regression is based on decision trees but presents linear models in the final nodes (leaves) of a decision tree instead of discrete values[45,46]. It also uses a boosting-like scheme to improve its prediction by building successive trees. For hyperparameters optimization, we used spatial cross-validation with 10 folds and grid search (more details in SI Note S4). The final model can be accessed at the online repository of this study along with sample codes of its use[44]. The model performance on the test set was assessed using the Kling-Gupta efficiency index[33] (KGE), coefficient of determination ($r^2$), mean absolute error (MAE), and percentage bias (PBIAS).

*Model interpretation*

The model interpretability gap was addressed using the Shapley values technique[47,48]. Shapley values is a model-agnostic technique based on game theory to fairly distribute the effect of the explanatory variables on the prediction[22]. For a single prediction, it returns the average expected marginal contribution of each explanatory variable in relation to the average prediction of the reference dataset. Here, we used the entire dataset as a reference and 1000 coalitions to obtain more accurate/stable shapley values. The shapley values were computed and aggregated for the entire dataset in order to obtain a fair understanding of the model mechanics.

*Model area of applicability*

To investigate the model applicability to new unseen data, we used the area of applicability (AoA) method[23]. This method consists of calculating the dissimilarity between new data and the model's known domain (training set). The dissimilarity index (DI) is based on the Euclidean distance between the standardized explanatory variables weighed by their importance for the ML model[23]. The method uses the model training cross-validation to investigate DI for unseen data and suggests a threshold to classify new data as inside or outside the model AoA. We also further investigate the relationship between average expected error and DI for the test set.

**Global extrapolation**

The ML model was used to estimate the green-blue WFP over the world (excluding Antarctica). The 17 explanatory variables were summarised for grid cells with a resolution of 0.25º x 0.25º and considering the recent period from 2000 to 2019. As the model was developed primarily to estimate R/P, we also derived the corresponding average green WFP as ET/P ≈ 1 – R/P. The predicted green-blue WFP were constrained to the range 0-1, as predictions outside the training domain (high DI) could fall outside this range. The associated global DI was also investigated.

## Supplemental information

Document S1. Figures S1–S10, Tables S1–S3, and Notes S1–S4.

## Acknowledgments

The authors would like to acknowledge the Bolin Centre for Climate Research at Stockholm University for the funding provided for this research (related to RA7 project SUB-ECOCLIM).

## Authors contributions

D.A. and G.D. contributed equally to study conceptualization, analysis of results, illustration, and writing of the paper. D.A. was the main responsible for compiling the data, modelling, and providing open access to the code and study data.

## Declarations of interests

The authors declare no competing interests

**Title:**

Supplemental Information to: "The global freshwater system: Patterns and predictability of green-blue flux partitioning"

**Authors:**

Daniel Althoff[1,*], Georgia Destouni[1,**]

**Affiliations:**

[1]Department of Physical Geography, Bolin Centre for Climate Research, Stockholm University, Stockholm, Sweden

*Correspondence: daniel.althoff@natgeo.su.se

**Correspondence: georgia.destouni@natgeo.su.se

# Contents

**Table S1.** Summary of data sources for hydrological catchments and daily runoff data.

| Abbrev. | Catchments | Period** | Mean obs. period (years) | Mean obs. avail. (%) |
|---|---|---|---|---|
| GRDC[1] | 1278 | 1980 – 2020 | 37.6 | 98.8 |
| CAMELS[2] | 622 | 1980 – 2014 | 34.2 | 99.7 |
| CAMELS-AUS[3] | 193 | 1980 – 2014 | 34.9 | 98.0 |
| CAMELS-BR[4,*] | 865 | 1980 – 2018 | 36.9 | 98.7 |
| CAMELS-CL[5] | 115 | 1980 – 2018 | 35.6 | 94.6 |
| CAMELS-GB[6] | 541 | 1980 – 2015 | 35.2 | 99.0 |

*Only selected catchments[4] were considered. **We only considered the period after 1980.

**Table S2.** Land cover classification system (LCCS) and classification adopted in this study, based on the land cover maps of the Climate Research Data Package (CRDP) generated within the European Space Agency (ESA) Climate Change Initiative – Land Cover (CCI-LC) project, covering the time period from 1992 to present.

| Value | LCCS | Reclass. |
|---|---|---|
| 0 | No data | No data |
| 10 | Cropland, rainfed | Agriculture |
| 11 | Herbaceous cover | Agriculture |
| 12 | Tree or shrub cover | Agriculture |
| 20 | Cropland, irrigated or post-flooding | Agriculture |
| 30 | Mosaic cropland (>50%) / natural vegetation (tree, shrub, herbaceous cover) (<50%) | Agriculture |
| 40 | Mosaic natural vegetation (tree, shrub, herbaceous cover) (>50%) / cropland (<50%) | Agriculture |
| 50 | Tree cover, broadleaved, evergreen, closed to open (>15%) | Forest |
| 60 | Tree cover, broadleaved, deciduous, closed to open (>15%) | Forest |
| 61 | Tree cover, broadleaved, deciduous, closed (>40%) | Forest |
| 62 | Tree cover, broadleaved, deciduous, open (15-40%) | Forest |
| 70 | Tree cover, needleleaved, evergreen, closed to open (>15%) | Forest |
| 71 | Tree cover, needleleaved, evergreen, closed (>40%) | Forest |
| 72 | Tree cover, needleleaved, evergreen, open (15-40%) | Forest |
| 80 | Tree cover, needleleaved, deciduous, closed to open (>15%) | Forest |
| 81 | Tree cover, needleleaved, deciduous, closed (>40%) | Forest |
| 82 | Tree cover, needleleaved, deciduous, open (15-40%) | Forest |
| 90 | Tree cover, mixed leaf type (broadleaved and needleleaved) | Forest |
| 100 | Mosaic tree and shrub (>50%) / herbaceous cover (<50%) | Forest |
| 110 | Mosaic herbaceous cover (>50%) / tree and shrub (<50%) | Grassland |
| 120 | Shrubland | Shrubland |
| 121 | Shrubland evergreen | Shrubland |
| 122 | Shrubland deciduous | Shrubland |
| 130 | Grassland | Grassland |
| 140 | Lichens and mosses | Sparse vegetation |
| 150 | Sparse vegetation (tree, shrub, herbaceous cover) (<15%) | Sparse vegetation |
| 151 | Sparse tree (<15%) | Sparse vegetation |
| 152 | Sparse shrub (<15%) | Sparse vegetation |
| 153 | Sparse herbaceous cover (<15%) | Sparse vegetation |
| 160 | Tree cover, flooded, fresh or brackish water | Forest |
| 170 | Tree cover, flooded, saline water | Forest |
| 180 | Shrub or herbaceous cover, flooded, fresh/saline/brakish water | Wetland |
| 190 | Urban areas | Settlement |
| 200 | Bare areas | Bare area |
| 201 | Consolidated bare areas | Bare area |
| 202 | Unconsolidated bare areas | Bare area |
| 210 | Water bodies | Water |
| 220 | Permanent snow and ice | Snow and ice |

**Table S3.** Summary of data sources for predictor variables.

| Abbrev. | Spatial res. | Temporal res. | Period | Description |
|---|---|---|---|---|
| MSWEP[7] | 0.10° | Daily | 1979 – present | Daily precipitation (global) |
| CPC/NOAA[8] | 0.25° | Daily | 1979 – present | Daily min. and max. surface temperature (global) |
| ESA CCI-LC[9,10] | 300 m | Annual | 1992 – 2020 | Land cover classification maps (global) |
| MERIT-DEM[11] | 90 m | - | - | Digital elevation model (90N-60S) |

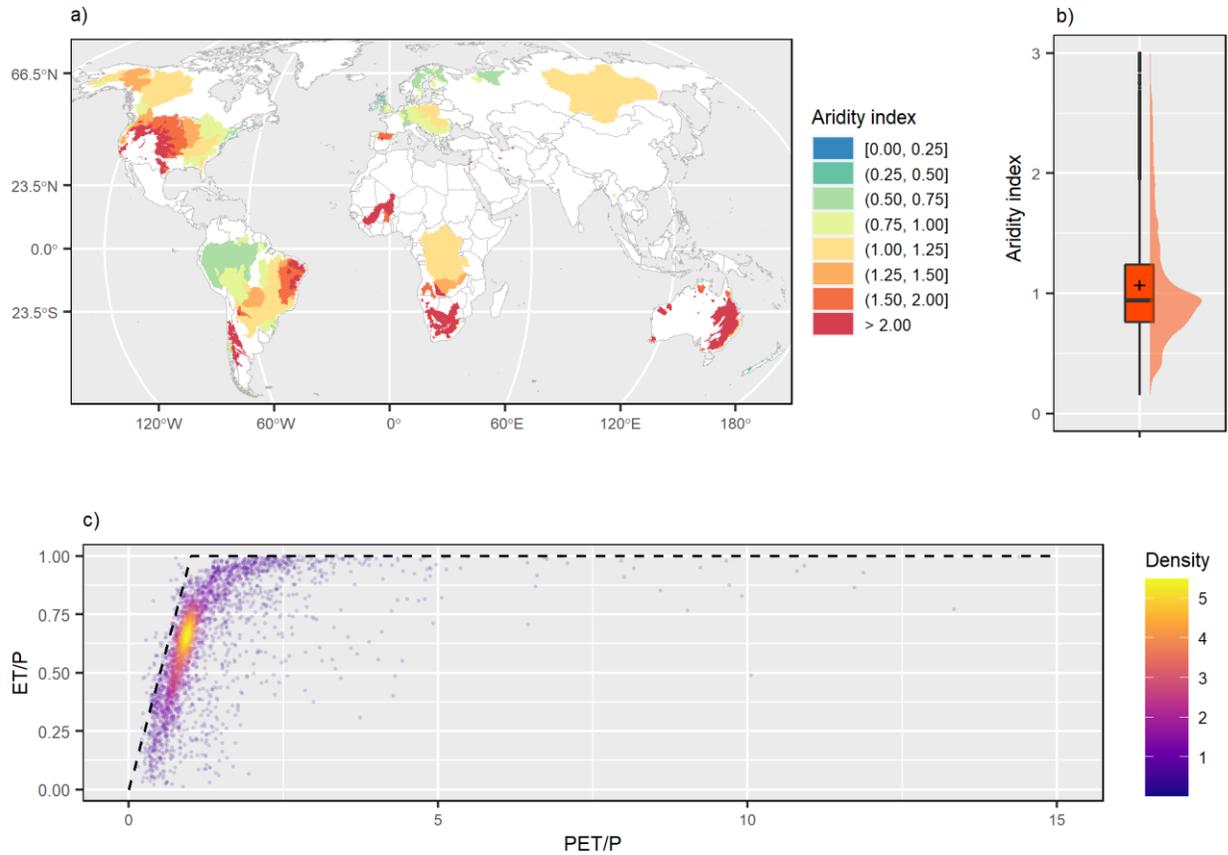

**Figure S1.** The study catchments' (a) aridity index PET/P; and the PET/P (b) statistical distribution and (c) Budyko space distribution (versus ET/P) across all catchments, where PET is average potential evapotranspiration, ET is average actual evapotranspiration, and P is average precipitation for each catchment over the study period.

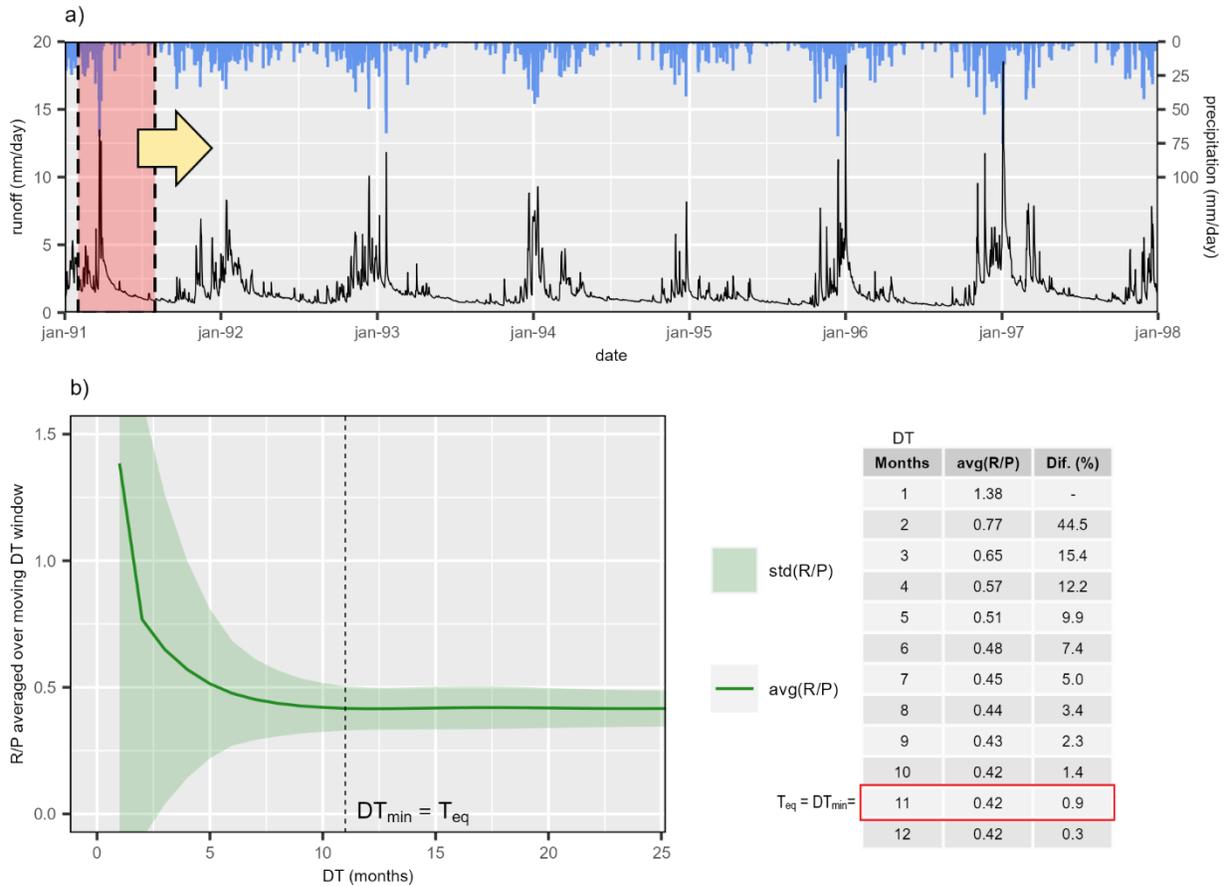

**Figure S2.** Schematic illustration of how water flux equilibration time ($T_{eq}$) is estimated for an arbitrary hydrological catchment. Panel (a) illustrates a moving time-window (DT) averaging of daily R/P (runoff/precipitation) for the example of DT = 6 months. Panel (b) illustrates average R/P (avg(R/P), green line) among the DT windows moving over the whole data time series, and the stabilization of avg(R/P) to a more or less constant value for increasing DT, along with the temporal standard deviation (std(R/P), green shade) around avg(R/P) for the moving DT windows. The flux equilibration time $T_{eq}$ is quantified as equal to the minimum DT value ($T_{eq} = DT_{min}$), for and beyond which avg(R/P) values obtained for any DT ≥ $DT_{min}$ differ less than 1% from each other.

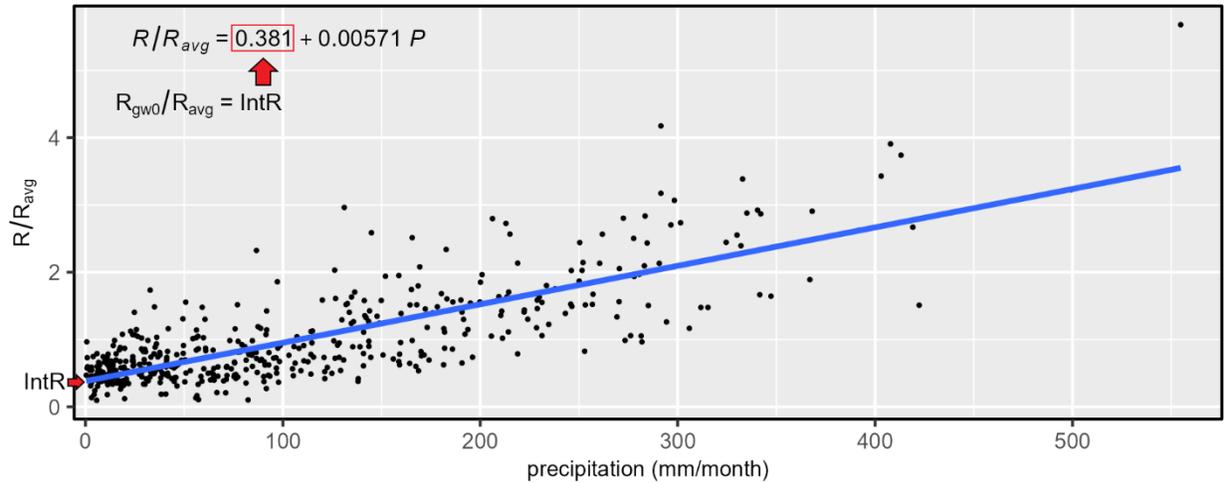

**Figure S3.** Schematic illustration of the approach to estimate average groundwater contribution to surface water runoff under zero monthly precipitation conditions ($R_{gw0}$) from the regression line (blue) fitted to the data points (black dots) of monthly runoff (R), normalized with average runoff over the whole runoff time series ($R_{avg}$), versus monthly precipitation (P) for each study catchment. The groundwater indicator variable $R_{gw0}/R_{avg}$ is quantified as equal to the regression line intercept (IntR) for $0 \leq IntR \leq 1$. Negative intercept, $IntR < 0$, implies no surface water runoff (essentially dry streams and/or stream networks with stillstanding water) for zero monthly P. This is not an unrealistic condition, but one for which groundwater flow cannot be simply estimated from the R vs P regression line; 1,274 (35% of all) study catchments with $IntR < 0$ were therefore excluded from the $R_{gw0}/R_{avg}$ analysis. Moreover, $IntR > 1$ implies unrealistically large average runoff under zero monthly precipitation, as this would be even larger than the total average runoff $R_{avg}$ over the whole precipitation and runoff time series; 105 (3% of all) study catchments with $IntR > 1$ were therefore also excluded from the $R_{gw0}/R_{avg}$ analysis.

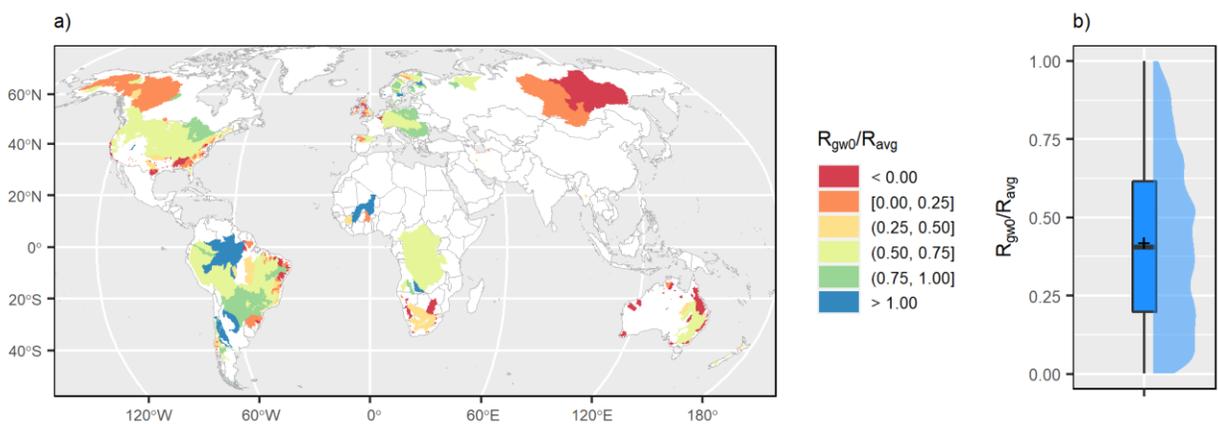

**Figure S4.** (a) Map of estimated average groundwater contribution to surface water runoff under zero monthly precipitation conditions ($R_{gw0}$) relative to total average surface water runoff ($R_{avg}$) for the investigated hydrological catchments, and (b) the statistical distribution of $R_{gw0}/R_{avg}$ among the catchments. See Figure S3 for explanation of how $R_{gw0}/R_{avg}$ is estimated.

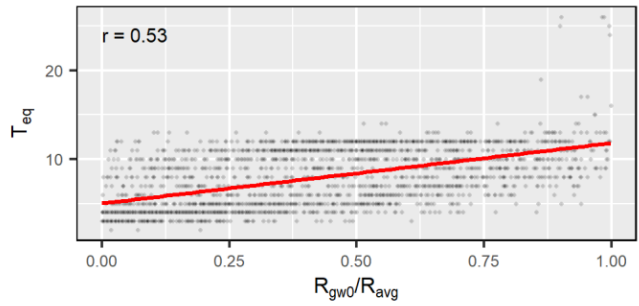

**Figure S5.** Correlation between flux equilibration time $T_{eq}$ (Figure S2, main Figure 1A-B) and monthly values (according to the definition for zero monthly precipitation) of groundwater indicator $R_{gw0}/R_{avg}$ (Figures S3-S4) among the study catchments.

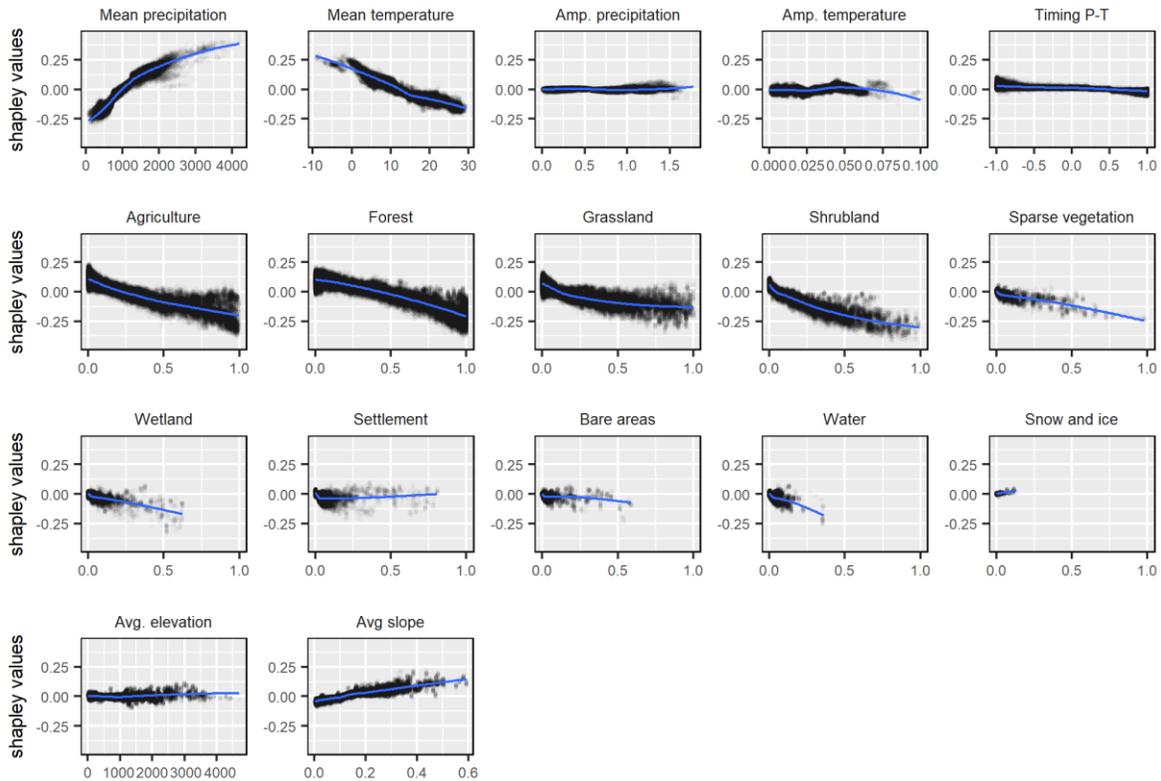

**Figure S6.** Marginal contributions (shapley values) for all explanatory variables of the ML model to dataset predictions of blue water flux partitioning (R/P). Shapley values relate to the average dataset prediction (R/P = 0.37, ET/P = 0.63). Positive [negative] shapley values indicate that explanatory variables are contributing towards increase in R/P [ET/P].

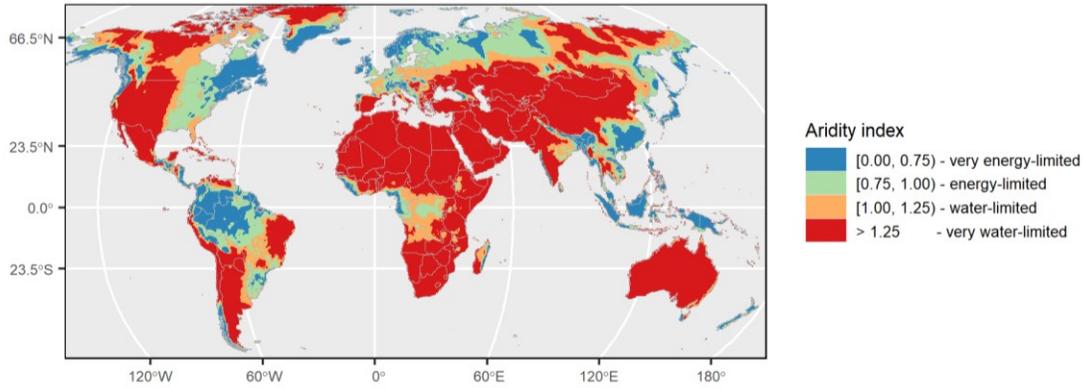

**Figure S7.** Pixel-wise aridity index around the world.

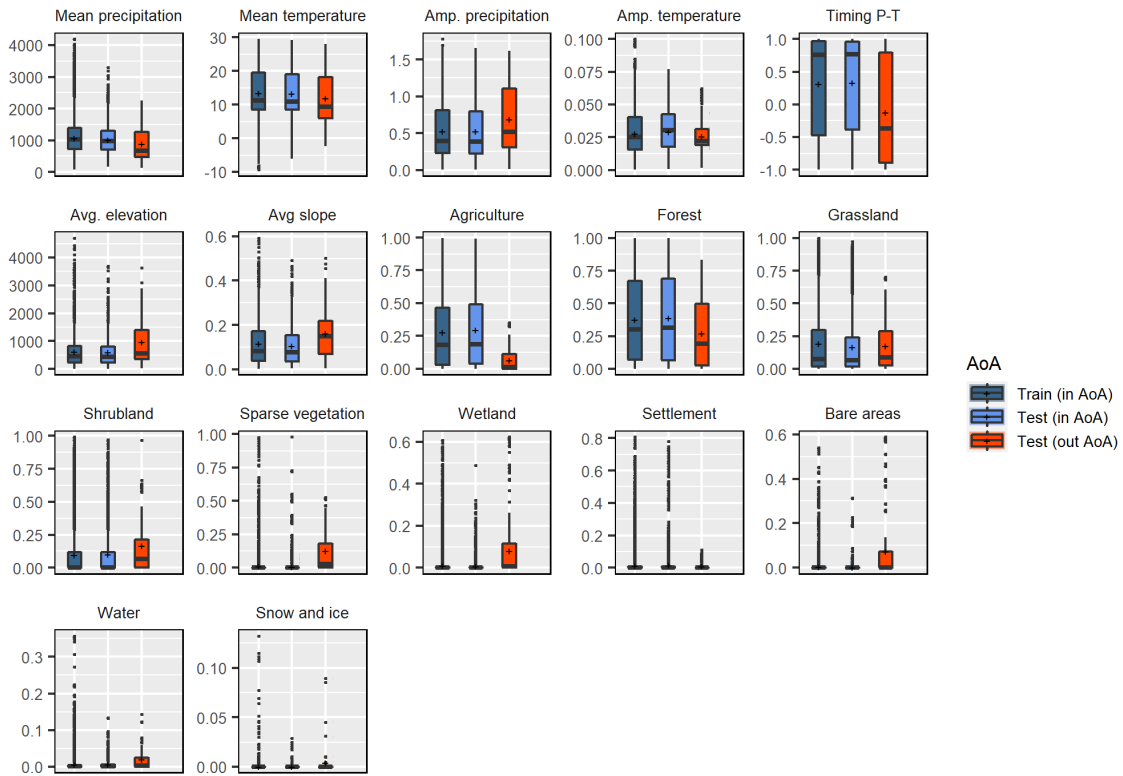

**Figure S8.** Statistical distribution of machine learning model explanatory variables for instances categorized as inside and outside the area of applicability (AoA). Here, we used the dissimilarity index limit of 0.27 to classify instances as inside or outside AoA.

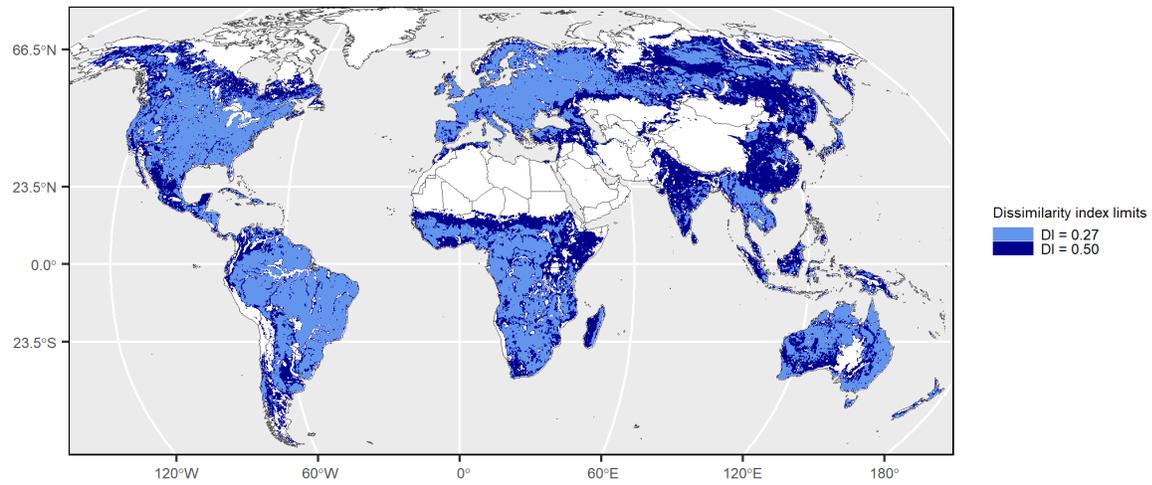

**Figure S9.** Global land areas inside the machine learning model area of applicability for different dissimilarity index limits.

*Catchments boundaries and daily runoff data.*

**Note S1**

The hydrological catchments (boundaries and runoff data) used in this study were compiled from the Global Runoff Database Centre (GRDC)[1] and multiples Catchment Attributes and Meteorology for Large-sample Studies (CAMELS) datasets, i.e., for the contiguous United States[2], Australia[3], Brazil[4], Chile[5], and Great Britain[6].

We discarded 14 catchments from the CAMELS-AUS dataset because they had notes concerning the accuracy of their boundaries delimitation. To avoid using catchments accounted for both the GRDC and CAMELS datasets, we also discarded catchments from the GRDC dataset when the distance between a catchment gauge station to the nearest CAMELS gauge station was shorter than the minimum distance between gauge stations in that CAMELS dataset. A total of 350 catchments were discarded.

5775 hydrological catchments had daily runoff records available between 1980 and 2019. We further discarded hydrological catchments with (i) less than 25 years of daily runoff data available since 1980 (1527 catchments) and (ii) less than 80% of data in all months (316 catchments), i.e., catchments which data consistently missing in the same period of the year. Because data points were summarised for every 5-year period of the time series (see Section S3), the catchments runoff series were also screened for at least one 5-year period with "complete" data, i.e., no more than 3 missing days in any given month of a year (42 catchments discarded). The runoff data was also screened to discard catchments with (iii) odd behaviours in time series, such as unreasonable shifts in average runoff or interference likely caused by human-operated infrastructures, e.g., dams (67 catchments), and (iv) series with the average annual runoff above the average annual precipitation (196 catchments discarded). Finally, because data availability of other explanatory variables were also considered, 22 more catchments were discarded. The final database has a total of 3614 hydrological catchments (Table S1, Figure S10).

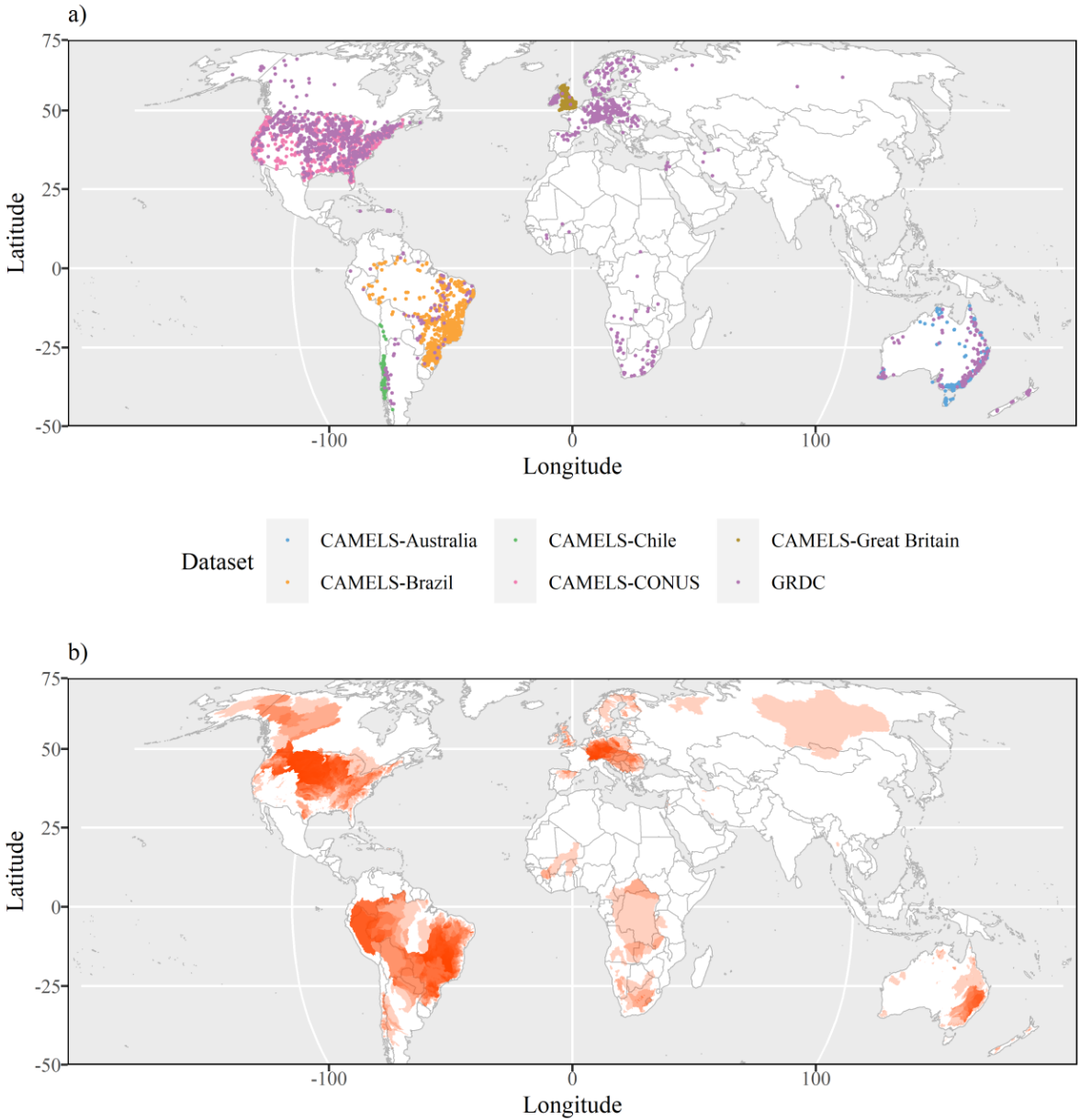

**Figure S10.** (a) Centroids and (b) spatial coverage of the hydrological catchments used in this study. Catchments polygons have a 75% transparency to help display regions with nested/overlapping catchments.

Spatial and tabular data wrangling was performed in R[12] using mostly the rgdal[13], terra[14], dplyr[15], tidyr[16], doParallel[17], and foreach[18] packages. For all visualizations, we used the ggplot2[19] package.

*Hydro-climatic, land-cover/use, and topographical indices*

**Note S2**

Catchments daily precipitation (P) time series were derived from the Multi-Source Weighted Ensemble Precipitation (MSWEP)[7] version 2.8. Catchments daily average temperature (T) time series were obtained by averaging the daily minimum and maximum temperature series derived from the Climate Prediction Center (CPC) Global Temperature data[8]. To obtain the catchment P and T annual averages, seasonal amplitudes, and phase lag between peak P and peak T over the year under a specific period, their respective time series were approximated to sine curves (Equations S1 and S2):

$$P = \bar{P}[1 + \delta_P \sin(2\pi(t - s_P)/\tau)] \quad (S1)$$

$$T = \bar{T} + \delta_T \sin(2\pi(t - s_T)/\tau) \quad (S2)$$

where P and T are the daily precipitation and temperature time series (mm, °C), respectively, $\bar{P}$ and $\bar{T}$ are the time-averaged precipitation and temperature (mm, °C), respectively, $\delta_P$ and $\delta_T$ are the precipitation and temperature seasonal amplitude (dimensionless, °C), respectively, $s_P$ and $s_T$ are the P and T phase shifts (days), respectively, $t$ is the time (day-of-year), and $\tau$ is the duration of the seasonal cycle (365 days). When approximating the sine curves, the variable time-averaged, seasonal amplitude, and phase shift are calibrated and must be positive. For the ML modelling, $\bar{P}$ and was scaled from daily to annual averages while the temperature seasonal amplitude, $\delta_T$, was normalized by the time-averaged temperature ($\delta_T^* = \delta_T/(\bar{T} + 273.15)$).

The phase lag between peak P and peak T (P-T timing) over the year can then be obtained from Eq. S3:

$$TI = \cos(2\pi(s_P - s_T)/365) \quad (S3)$$

where TI is the phase lag between peak P and peak T over the year, or timing index. TI ranges from -1 and 1, where 1 indicate that P and T both peak in the same moment, i.e., precipitation peaks during summer, 0 indicates a phase lag of 3 months, and -1 indicates the maximum phase lag of 6 months.

**Note S3**

The land-cover/use classes were derived from land cover maps of the Climate Research Data Package (CRDP). The CRDP was generated within the European Space Agency (ESA) Climate Change Initiative – Land Cover (CCI-LC) project[9,10] and covers the period from 1992 to present. The land cover classification system describes 22 main classes (level 1) and a total of 38 sub-classes (level 2). For simplicity, the classes were grouped into 10 broad categories (Table S2) that better correspond to the Intergovernmental Panel on Climate Change (IPCC) land categories[9]: agriculture, forest, grassland, shrubland, sparse vegetation, wetland, bare area, settlement, water, and permanent snow and ice.

The catchment mean elevation (m) and mean slope (m/m) were derived from the Multi-Error-Removed Improved-Terrain Digital Elevation Model (MERIT DEM)[11].

*Machine learning model training*

**Note S4**

The machine learning model was developed using the cubist regression[20–22]. The cubist regression hyperparameters "committees" and "neighbours" refer to the number of trees created in sequence, similar to boosting, and the number of nearest-neighbours points from the training set that can be used to adjust the final prediction, respectively.

Hyperparameter tuning was performed during training using a grid search (committees [0-30], neighbours [0-9]) and spatial cross-validation with 10 folds, i.e., data points from the same catchment were all allocated in the same fold.

We used the cubist[23] and caret[24] packages in R[12] for training the model and hyperparameter-tuning, and the CAST[25] package for the spatial partitioning of the training set for spatial cross-validation.